\documentclass[%
reprint,
superscriptaddress,
 amsmath,amssymb,
 aps,
pra,
longbibliography,
floatfix,
]{revtex4-2}
\usepackage{graphicx}
\usepackage{color}
\usepackage{hyperref}
\usepackage{float}
\usepackage{array, multirow}
\usepackage{amsmath}
\usepackage{newtxtext}              \usepackage{booktabs}
\usepackage{xcolor}
\usepackage{soul}
\usepackage{url}
\usepackage{qcircuit}
\usepackage[margin=0.8in]{geometry}

\hypersetup{colorlinks,
    linkcolor={blue!75!black!80!yellow},
    citecolor={blue!75!black!80!yellow},
    urlcolor={blue!75!black!80!yellow}
    }

\begin{document}

\title{Quantum State Preparation and Non-Unitary Evolution with Diagonal Operators}

\author{Anthony W. Schlimgen}
\affiliation{Department of Chemistry and The James Franck Institute, The University of Chicago, Chicago, IL 60637 USA}
\author{Kade Head-Marsden}
\affiliation{John A. Paulson School of Engineering and Applied Sciences, Harvard
University, Cambridge, MA 02138, USA}
\author{LeeAnn M. Sager-Smith}
\affiliation{Department of Chemistry and The James Franck Institute, The University of Chicago, Chicago, IL 60637 USA}
\author{Prineha Narang}
\affiliation{John A. Paulson School of Engineering and Applied Sciences, Harvard
University, Cambridge, MA 02138, USA}
\author{David A. Mazziotti}
\email{damazz@uchicago.edu}
\affiliation{Department of Chemistry and The James Franck Institute, The University of Chicago, Chicago, IL 60637 USA}

\date{Submitted May 5, 2022}

\begin{abstract}
Realizing non-unitary transformations on unitary-gate based quantum devices is critically important for simulating a variety of physical problems including open quantum systems and subnormalized quantum states.  We present a dilation based algorithm to simulate non-unitary operations using probabilistic quantum computing with only one ancilla qubit. We utilize the singular-value decomposition (SVD) to decompose any general quantum operator into a product of two unitary operators and a diagonal non-unitary operator, which we show can be implemented by a diagonal unitary operator in a 1-qubit dilated space. While dilation techniques increase the number of qubits in the calculation, and thus the gate complexity, our algorithm limits the operations required in the dilated space to a diagonal unitary operator, which has known circuit decompositions. We use this algorithm to prepare random sub-normalized two-level states on a quantum device with high fidelity. Furthermore, we present the accurate non-unitary dynamics of two-level open quantum systems in a dephasing channel and an amplitude damping channel computed on a quantum device. The algorithm presented will be most useful for implementing general non-unitary operations when the SVD can be readily computed, which is the case with most operators in the noisy intermediate-scale quantum computing era.
\end{abstract}

\maketitle

\section{Introduction}

Recent advances in quantum computation have enabled algorithm implementation on real quantum devices in the Noisy Intermediate-Scale Quantum (NISQ) regime~\cite{Preskill2018}. This regime is defined by low-qubit counts where decoherence times are relatively short and two-qubit gate errors remain problematic. The noise experienced by the device presents a challenge towards practical algorithm implementation that has led to a plethora of research across chemistry, physics, and engineering~\cite{Kais2014, Cao2019, HeadMarsdenFlick2020, McArdle2020, Motta2021, Bharti2022}. One area that has been gaining recent attention is algorithm development for the non-unitary time evolution of quantum systems.  Current quantum devices are typically unitary-gate-based, so non-unitary operators must be cast as unitary in order to be practically implementable. There are a variety of algorithms which have been developed to bypass this obstacle, including explicit mathematical dilations~\cite{Sweke2014, Sweke2015, Hu2020, HeadMarsden2021, Hu2021, Gaikwad2022, Schlimgen2021PRL}, quantum imaginary time evolution~\cite{Kamakari2022}, duality~\cite{Wei2016, Zheng2021}, the variational principal~\cite{Endo2020}, collision models~\cite{Cattaneo2022}, analog simulation~\cite{Kim2022}, and others~\cite{Sweke2016, Chenu2017, Su2020, Patsch2020, Garcia-Perez2020, Rost2020, Rost2021, Metcalf2022, Guimaraes2022, Xin:2017wl, Childs2017, Cleve2016}. The majority of these algorithms rely on some form of dilation, either mapping the operator to a larger Hilbert space, or adding ancilla qubits. Another way to view this problem is through the lens of non-normalized state preparation. Non-unitary operations are not norm conserving, so evolution of a state non-unitarily will result in a sub-normalized state.

Here, we present and demonstrate a dilation-based algorithm using non-unitary diagonal operators. Diagonal operators are relatively sparse, and have known circuit decompositions, which make them attractive for multi-qubit calculations. We show that non-unitary diagonals can be transformed to unitary diagonals with a one-qubit dilation. The algorithm is probabilistic, but the success probability of preparing the desired state can be improved with standard amplitude amplification techniques, because the desired state is known. Finally, the exact preparation of the desired state requires $\mathcal{O}(2^{k+1
})$ one- and two-qubit gates, where $k$ is the number of system qubits. Importantly, diagonal gates can also be implemented approximately with controlled error, requiring only a polynomial number of gates with respect to the number of qubits~\cite{Welch:2014}.

This dilation can be used for probabilistic state preparation as well as non-unitary evolution of quantum states. Dilated diagonals can be utilized to prepare both normalized and sub-normalized states. While sub-normalized states can always be normalized before a quantum simulation, the situation often arises where the state needs to be manipulated further on a quantum device in its sub-normalized form as, for example, in preparing linear combinations with other un-normalized states. In this work, we prepare a random selection of sub-normalized states on an IBM quantum device and perform a tomography of those states, demonstrating that we can achieve high-fidelity state preparation.

Sub-normalized states also arise in the context of non-unitary dynamics. Using the singular-value decomposition (SVD), we show that any non-unitary operator can be written in terms of two unitaries and a non-unitary diagonal operator, which can be dilated to a unitary. The classical cost of the algorithm is the cost of the SVD, which scales as $\mathcal{O}(r^3)$, where $r$ is the size of the original operator. Our algorithm requires only one ancilla qubit for any size operator, and the entangling operations between the state and ancilla are reduced to a diagonal gate, which is efficient to implement. In the limit of a large number of qubits, our dilated algorithm is approximately double the cost of a unitary propagation on the original Hilbert space in terms of circuit depth. On the other hand, compared to a general dilated unitary operator, our algorithm reduces the circuit depth by approximately half.

We outline the non-unitary diagonal operator implementation for state preparation in Section~\ref{theory1} and the SVD algorithm for non-unitary evolution in Section~\ref{theory2}. The computational methodology is laid out in Section~\ref{methods} then demonstrated on the preparation of sub-normalized states and open system dynamics in Sections~\ref{res1} and \ref{res2}, respectively.

\section{Theory}

\subsection{Non-Unitary Diagonal Operators and Quantum State Preparation}
\label{theory1}

Due to their sparsity, diagonal operators are attractive transformations for quantum simulation, and there are known algorithms for efficient implementation of unitary diagonals~\cite{Shende2006,Welch:2014}. Here we show that non-unitary diagonals can be implemented as unitary diagonal gates with only one ancilla qubit. Consider a non-unitary diagonal operator, $\hat{\Sigma}$ with complex entries $\sigma_{ii}$ on the diagonal, and assume the magnitude of each element is less than or equal to unity. We can directly construct a unitary diagonal operator,
\begin{equation}
    \hat{U}_{\hat{\Sigma}} = \begin{pmatrix} \hat{\Sigma}_+ & 0 \\ 0 & \hat{\Sigma}_-
    \end{pmatrix},
\label{eq:szDiag}
\end{equation}
where,
\begin{equation}
    \hat{\Sigma}_{ii\pm} =  \sigma_{ii} \pm i\sqrt{\frac{1-\lvert\lvert \sigma_{ii} \rvert\rvert^2}{\lvert\lvert \sigma_{ii} \rvert\rvert^2}}\sigma_{ii}.
    \label{eq:singSz}
\end{equation}
We also write Eq.~\ref{eq:szDiag} as $\hat{U}_{\hat{\Sigma}} = \hat{\Sigma}_{+} \oplus \hat{\Sigma}_{-}$ where $\oplus$ is the block, or direct sum operator. We emphasize that $\hat \Sigma = \hat \Sigma_+ + \hat \Sigma_-$. If the size of the non-unitary diagonal operator is $r$, then the size of the dilated unitary $\hat{U}_{\hat{\Sigma}}$ is $2r$. This implies that the dilated unitary can be implemented on a quantum device with only one ancilla qubit, because we only need to double the size of original Hilbert space. Importantly, the dilated unitary is trivial to compute from the non-unitary diagonal, requires no matrix operations to generate the unitary, and requires no recursive decomposition to compute rotation angles.

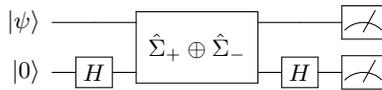
\begin{figure}[ht!]
$$
\begin{array}{c}
\Qcircuit @C=1em @R=.7em {& \lstick{|\psi\rangle} & \qw & \multigate{1}{\hat{\Sigma}_+ \oplus \hat{\Sigma}_-}  & \qw & \meter \\& \lstick{|0\rangle} & \gate{H} & \ghost{\hat{\Sigma}_+ \oplus \hat{\Sigma}_-}  & \gate{H} &  \meter}
\end{array}
$$
\caption{Circuit for implementing a non-unitary diagonal operator, $\hat \Sigma$, on $|\psi\rangle$ using only unitary gates with a one-qubit dilation. The final gate on each rail is the measurement gate, $H$ is the Hadamard gate, and the multi-qubit gate is the diagonal operator in Eq.~\ref{eq:szDiag}. The algorithm is successful when the ancilla is measured in state $|0\rangle$.}
\label{fig:circ0}
\end{figure}

We can achieve the probabilistic application of the non-unitary diagonal operator by preparing the ancilla qubit in the superposed state, and recombining the states after application of $\hat{U}_{\hat{\Sigma}}$, as shown in Figure ~\ref{fig:circ0}.
The algorithm is successful when the ancilla qubit is in state $|0\rangle$, and the non-unitary $\hat{\Sigma}$ is applied to the state $|\psi\rangle$. The desired state is known, so the success probability can be increased with standard techniques of amplitude amplification~\cite{Brassard2002}.

It is straightforward to apply this technique to quantum state preparation. Suppose we wish to prepare a quantum state $|\phi\rangle$, the diagonal operator generated from $|\phi\rangle$ on the diagonal is in general non-unitary, but can be applied using the procedure above. If we initialize the system in the mixed state by applying a Hadamard gate, $H$, to each qubit, then the algorithm will result in the desired state preparation probabilistically,
\begin{equation}
    \lvert \Psi \rangle =   (I^{\otimes k} \otimes H) D^{k+1} (H^{\otimes k}\lvert 0\rangle^{\otimes k} \otimes H\lvert 0\rangle).
\end{equation}


Here, $D^{k+1}$ is a diagonal operator in the form of Eq.~\ref{eq:szDiag} which spans the system and ancilla qubits, $\vert\Psi\rangle$ is the total wavefunction of the system and ancilla qubits, and the desired state $\lvert\phi\rangle$ will be prepared when the ancilla is in state $\lvert 0\rangle$. This has some similarity to standard state initialization techniques, in that both utilize a diagonal operator. Standard techniques, however, compute the unitary required to take the ground state to the desired state. Instead, we take the superposed state to the desired state probabilistically, through a non-unitary transformation with an ancilla qubit.

This approach can be used for preparation of sub-normalized states as well as normalized states. An ensemble state is characterized by a positive sum of density matrices,
\begin{equation}
\rho = \sum_i p_i \rho_i.
\end{equation}
While each $\rho_i$ is a pure state and can be individually normalized, it can be useful to prepare the sub-normalized states $p_i \rho_i$, which can be further transformed on a quantum device. Probabilistic normalized state preparation in this form is thus more expensive in terms of gate depth compared to standard state initialization techniques due to the ancilla qubit and larger diagonal operator. In contrast, the computation of the diagonal operator needed for this state preparation is immediately available by the simple arithmetic in Eq.~\ref{eq:singSz} and requires no recursive decomposition. Furthermore, our probabilistic approach allows for the preparation of non-normalized states, and therefore the implementation of non-unitary operators. We discuss the circuit complexity for diagonal operators in the next section.

\subsection{Non-Unitary Evolution}
\label{theory2}

A quantum system represented by a density matrix, $\rho$, undergoing non-unitary evolution can be characterized by unitary evolution in a dilated Hilbert space, where the state is coupled to some environment, $\sigma_B$. This can be expressed with Stinespring's dilation, where the propagated density matrix, $\rho(t)$, is given by the partial trace of the unitary evolution of the interacting system and environment~\cite{Stinespring1955, Buscemi2003, Nielsen2010, Sweke2015}.

A more cost-effective method is the Sz.-Nagy dilation, which was originally applied to contraction mappings and requires only one ancilla qubit for implementation~\cite{Hu2020,Hu2021}. The one-dilation for a non-unitary operator $\hat{M}$ can be written as,
\begin{equation}
    \hat{U}_{\hat{M}} = \begin{pmatrix}
    \hat{M} & \sqrt{I - \hat{M}\hat{M}^\dagger}\\
    \sqrt{I - \hat{M}^\dagger \hat{M}} & -\hat{M}^\dagger \\
    \end{pmatrix},
    \label{eq:sznagy}
\end{equation}
where the off-diagonal elements are known as the defect operators. The dilation is guaranteed when $\hat{M}$ is a contraction, so the square root in the defect operator is defined, and any bounded operator can be shifted to a contraction~\cite{Hu2020}.
While this dilation requires only one ancilla qubit, the dilated unitary acts over the entire dilated space, which can lead to quantum circuits with high gate counts when the system becomes large.

Consider instead the singular value decomposition (SVD) for $\hat M$,
\begin{equation}
    \hat M = \hat U \hat\Sigma \hat V^\dagger,
\end{equation}
where $\hat U$ and $\hat V^\dagger$ are unitary operators, and $\hat \Sigma$ is a diagonal operator of the singular values of $\hat M$, which are always real and non-negative. Since $\hat U$ and $\hat V^\dagger$ are unitary, they can be implemented on the space that is the same size as the original operator $\hat{M}$, but $\hat{\Sigma}$ is non-unitary and must be dilated, as shown above. The SVD has been used to analyze non-unitary operations in other quantum algorithms~\cite{Williams:2004up,Liu:2021tz}; however, we show here that the singular value matrix can be implemented efficiently as a diagonal operator, and is the only operator that must span the dilated space. Importantly, the implementation of the non-unitary operation here does not depend on an expansion parameter as in other works~\cite{Williams:2004up,Liu:2021tz,Schlimgen2021PRL}, so the operation can be implemented exactly, so long as decompositions for $U$ and $V^\dagger$ can be performed.

The formulation of the singular values in Eq.~\ref{eq:singSz} is inspired by the the Sz. Nagy dilation in Eq.~\ref{eq:sznagy}, where the square root term is reminiscent of the defect operator. Indeed the unitary in Eq.~\ref{eq:szDiag} is related to the standard form of the Sz. Nagy dilation by a rotation; however, the present formulation maintains the diagonal structure of the singular value matrix, which generally results in shallower circuits.

While the singular values are always real and non-negative, Eq.~\ref{eq:singSz} is suitable for any real or complex number so long as its magnitude is bound by one. In the case of singular values of magnitude greater than one, simply dividing $\hat{\Sigma}$ by the largest singular value will ensure that Eq.~\ref{eq:szDiag} remains unitary. In fact, dividing by the largest singular value results in an operator that is always a contraction, which guarantees an Sz. Nagy dilation. This observation is related to that of Hu \textit{et al.} who showed that since every bounded operator is bounded by its Hilbert-Schmidt norm, it can be shifted to an operator that is always a contraction~\cite{Hu2020}. Shifting the operator, or scaling it by the largest singular value, will effect the success probability of the algorithm, but the success probability can be driven towards unity with amplitude amplification.

Generating the singular values requires computing the sum of $\hat \Sigma_+$ and $\hat \Sigma_-$. Linear combinations of unitary operators are in general non-unitary; however, we can implement linear combinations of unitary operators using the block sum, or direct sum, approach, as in Eq.~\ref{eq:szDiag}. The non-unitary propagation of $\lvert\psi\rangle$, including the linear combinations, can be performed with the following transformation,
\begin{equation}
\begin{aligned}
    &\begin{pmatrix} I & I \\ I & -I \end{pmatrix} \begin{pmatrix} \hat U & 0 \\ 0 & \hat U \end{pmatrix}
    \begin{pmatrix} \hat{\Sigma}_+  & 0 \\ 0 & \hat{\Sigma}_- \end{pmatrix}
    \begin{pmatrix} \hat V^\dagger & 0 \\ 0 & \hat V^\dagger \end{pmatrix}
    \begin{pmatrix} |\psi \rangle \\ |\psi \rangle \end{pmatrix} \\
    &=\begin{pmatrix}
     \hat U (\hat{\Sigma}_+ + \hat{\Sigma}_-) \hat V^\dagger |\psi \rangle \\ \hat U (\hat{\Sigma}_+ - \hat{\Sigma}_-) \hat V^\dagger |\psi \rangle \end{pmatrix}
     =  \begin{pmatrix}
     \hat M |\psi \rangle \\ \hat{M}^-|\psi \rangle \end{pmatrix},
\end{aligned}
\label{eq:mat_u}
\end{equation}
where we note that $\hat{M}=\hat U (\hat{\Sigma}_+ + \hat{\Sigma}_-) \hat V^\dagger$, and define $\hat{M}^-=\hat U (\hat{\Sigma}_+ - \hat{\Sigma}_-) \hat V^\dagger$. The linear combinations are achieved through the final Hadamard gate on the ancilla qubit, as discussed elsewhere~\cite{Schlimgen2021PRL, Xin:2020wh,Berry2014a,Childs2012}. Finally, while we have shown this transformation for a pure state, using the eigendecomposition of a mixed state along with linearity allows for the straightforward application of Eq.~\ref{eq:mat_u} to mixed states. We show in the next section that, after assuming the classical cost of the SVD, the decomposition in Eq. ~\ref{eq:mat_u} can be implemented on a quantum circuit, with minimal operations spanning the entire dilated space. \\

\section{Methods}
\label{methods}

\begin{figure}[ht!]
$$
\begin{array}{c}
\Qcircuit @C=1em @R=.7em {& \lstick{|\psi\rangle} & \gate{\hat{V}^\dagger}  & \multigate{1}{\hat{\Sigma}_+ \oplus \hat{\Sigma}_-}  & \gate{\hat{U}} & \meter \\& \lstick{|0\rangle} & \gate{H} & \ghost{\hat{\Sigma}_+ \oplus \hat{\Sigma}_-}  & \gate{H} &  \meter}
\end{array}
$$
\caption{Circuit for implementing the non-unitary operator $\hat M$ on $|\psi\rangle$ using only unitary gates with a one-qubit dilation. The final gate on each rail is the measurement gate, and all other gates are defined in the text. The only operator acting over the dilated space is the diagonal operator, and the linear combinations are performed using the Hadamard gates.}
\label{fig:circ}
\end{figure}
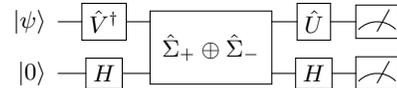

 We show the general circuit for implementing Eq.~\ref{eq:mat_u} in Fig.~\ref{fig:circ}. The non-unitary propagation of a wavefunction of length $r$, can be implemented on a quantum circuit with $k+1$ qubits, where $k \ge \log_2(r)$. This circuit requires the implementation of two $k$-qubit operators, $\hat V^\dagger$ and $\hat U$, along with a $(k+1)$-qubit diagonal operator, $\hat{\Sigma}_+ \oplus \hat{\Sigma}_-$. This formulation leads to reduction in algorithm complexity because the only operator to act on the larger one-dilated space is a diagonal operator, which is in general less expensive to implement, when compared to a dense one-dilated operator.

To see this reduction in cost, let the dilated space require $d=k+1$ qubits, and consider a general unitary that can be decomposed into $\mathcal{O}(d^2 2^{2d})$ one-qubit and CNOT gates~\cite{Nielsen2010}. The diagonal gate operating over $d$ qubits can be exactly decomposed into at most ${2^{d+1} - 3}$ fundamental gates~\cite{Bullock:2004}. and the operators $\hat U$ and $\hat V^\dagger$ can each be implemented with $\mathcal{O}((d-1)^2 2^{2d-2})$ gates. To leading order, therefore, our algorithm reduces the circuit depth by factor of two, resulting in a complexity of $\mathcal{O}(d^22^{2d-1})$, compared to the dense $d$-qubit operator.

As $d$ becomes large, the cost of our algorithm is dominated by the implementation of the unitaries $\hat{U}$ and $\hat{V}^\dagger$, and the cost of the dilated diagonal operator becomes negligible. Seen in this way, our algorithm to implement a $k$-qubit non-unitary operator is twice as expensive as a $k$-qubit unitary operator, in the large-$k$ limit. The exact implementation of diagonal gates results in the gate counts above, however, approximate polynomial implementations of diagonal gates are also known~\cite{Welch:2014}.

\section{Results}

\subsection{Preparation of sub-normalized states}
\label{res1}

Here we show the algorithm's utility in preparing sub-normalized quantum states on quantum devices. We randomly generated 98 complex sub-normalized one-qubit states, $|\psi_1\rangle$, from normalized two-qubit states, $|\psi_2\rangle$,
\begin{equation}
\begin{aligned}
    |\psi_2\rangle &= a_{00} |00\rangle + a_{01} |01\rangle + a_{10} |10\rangle + a_{11} |11\rangle \\
    |\psi_1\rangle &= a_{00} |0\rangle + a_{01} |1\rangle.
    \end{aligned}
\end{equation}
The state $|\psi_1\rangle$ is then prepared probabilistically using Eq.~\ref{eq:singSz} and the circuit in Fig.~\ref{fig:circ0}. The random states were generated with a seeded random-number generator in python. The average norm of the states is $0.67 \pm 0.12$.

 For each state we prepared a two-qubit circuit, initializing each qubit with a Hadamard gate, followed by implementing the diagonal operator in Eq.~\ref{eq:singSz}, using the diagonal operator decomposition available in Qiskit~\cite{Qiskit,Bullock:2004}. We perform full tomography of the 1-RDM which requires 3 circuits for each state, $|\psi_1\rangle$, then compute the fidelity and distance between the exact classical state, $\rho_E$, and the state from device tomography, $\rho_S$~\cite{Nielsen2010}. The fidelity is given by
\begin{equation}
    F(\rho_S, \rho_E) = \textrm{Tr}\bigg(\sqrt{\sqrt{\rho_S}\rho_E\sqrt{\rho_S}}\bigg)^2,
\end{equation}
and the distance is the Frobenius norm of the difference of the exact and simulated density matrices. The distance is computed between the un-normalized states, while the fidelity is computed with respect to the normalized density matrices.

Table~\ref{tab:stateprep} shows the accuracy of the state preparation for varying shot counts on the IBM Lagos device, utilizing error-mitigation techniques available in Qiskit \cite{lagos}. With $2^6$ samples, the fidelity is somewhat poor at 0.93; however, with only $2^{10}$ samples the fidelity of the prepared states has converged to be nearly exact at 0.99. The distance measure reveals the same trends in accuracy.
\begin{table}[ht]
    \centering
    \begin{tabular}{c c c}
    \hline \hline
    Samples & Distance & Fidelity \\
    \hline
        \vrule width 0pt height 2.2ex
    $2^6$ & 0.17 $\pm$ 0.07 & 0.93 $\pm$ 0.06 \\
    $2^8$ & 0.09 $\pm$ 0.04 & 0.98 $\pm$ 0.02 \\
    $2^{10}$ & 0.06 $\pm$ 0.02 & 0.99 $\pm$ 0.01 \\
    $2^{12}$ & 0.06 $\pm$ 0.02 & 0.99 $\pm$ 0.01 \\
    $2^{14}$ & 0.06 $\pm$ 0.02 & 0.99 $\pm$ 0.01 \\
    \hline \hline
    \end{tabular}
        \caption{Average error measures $\pm$ standard deviation for simulated sub-normalized one-qubit states for increasing simulation sampling (shots), where the distance and the fidelity are defined in the text. Results averaged over 98 randomly generated sub-normalized states prepared on the quantum device IBMQ Lagos with error mitigation.}
    \label{tab:stateprep}
\end{table}

Preparing these sub-normalized state probabilistically provides the groundwork for exploring non-unitary operations on unitary-gate-based quantum devices. Initializing the system as a superposition immediately provides the form of the diagonal unitary required to prepare the state, without needed recursive decomposition of rotation angles. Using this algorithm, we have prepared known sub-normalized states which would be the result of non-unitary operations. In the next section, we demonstrate how to extend this algorithm to performing the non-unitary action of a general operator on a unitary-gate-based architecture.

\subsection{Non-unitary evolution}
\label{res2}

The time evolution of an open quantum system can be described by the operator-sum formulation in terms of the Kraus operators,
\begin{equation}
    \rho(t) = \sum_i \hat K_i \rho(0) \hat K_i^\dagger,
\end{equation}
where the $\hat K_i$'s are the Kraus maps, and $\rho(t)$ is the density matrix at time $t$. In general, the Kraus maps are non-unitary; however, they are always contraction mappings,
\begin{equation}
   \sum_i \hat K_i \hat K_i^\dagger \leq I.
\end{equation}
Because the Kraus operators are always contraction mappings, their singular values are always bounded above by one, which makes Eq.~\ref{eq:singSz} directly applicable without rescaling of the singular values. One can decompose the $\hat K_i$ using the singular-value decomposition and implement the operator as described above, as two unitaries on the original Hilbert space and a unitary diagonal on the dilated space.

Here, we simulate two single-qubit systems, whose Kraus operators are either in diagonal or near-diagonal form, so the SVD is readily computed analytically. We used IBM's Lagos device to simulate the dynamics, and perform full tomography of the system density matrix. Each simulation was performed once with the device maximum 32000 samples, as well as in-built error mitigation protocols, in Qiskit. We present the error mitigated data for all results. Each circuit requires two qubits, one system qubit and one ancilla qubit.

For an initial example we simulate a two-level dephasing channel, with a single qubit coupled to a one-qubit bath with a $Z\otimes Z$ interaction. This results in diagonal Kraus operators,
\begin{equation}
    \begin{aligned}
    \hat K_0 &= \sqrt{\lambda_0} \begin{pmatrix} e^{i\theta t} & 0 \\ 0 & e^{-i\theta t} \end{pmatrix} \\
    \hat K_1 &= \sqrt{\lambda_1} \begin{pmatrix} e^{-i\theta t} & 0 \\ 0 & e^{i\theta t} \end{pmatrix},
    \end{aligned}
    \label{eq:phaseConst}
\end{equation}
where $\lambda_0 + \lambda_1 = 1$.  Here we choose the dephasing angle $\theta=0.5$ ps$^{-1}$, with $\lambda_0 = 0.7$ and $\lambda_1 = 0.3$. It is clear from Eq.~\ref{eq:phaseConst} that $\hat K_i/\sqrt{\lambda_i}$ is unitary and could be rescaled by $\lambda_i$ classically; however, here we simulate the non-unitary operators $\hat K_i$ on the device to show the non-unitary time propagation on a unitary-gate-based architecture.

\begin{figure}
    \centering
    \includegraphics[width=\columnwidth]{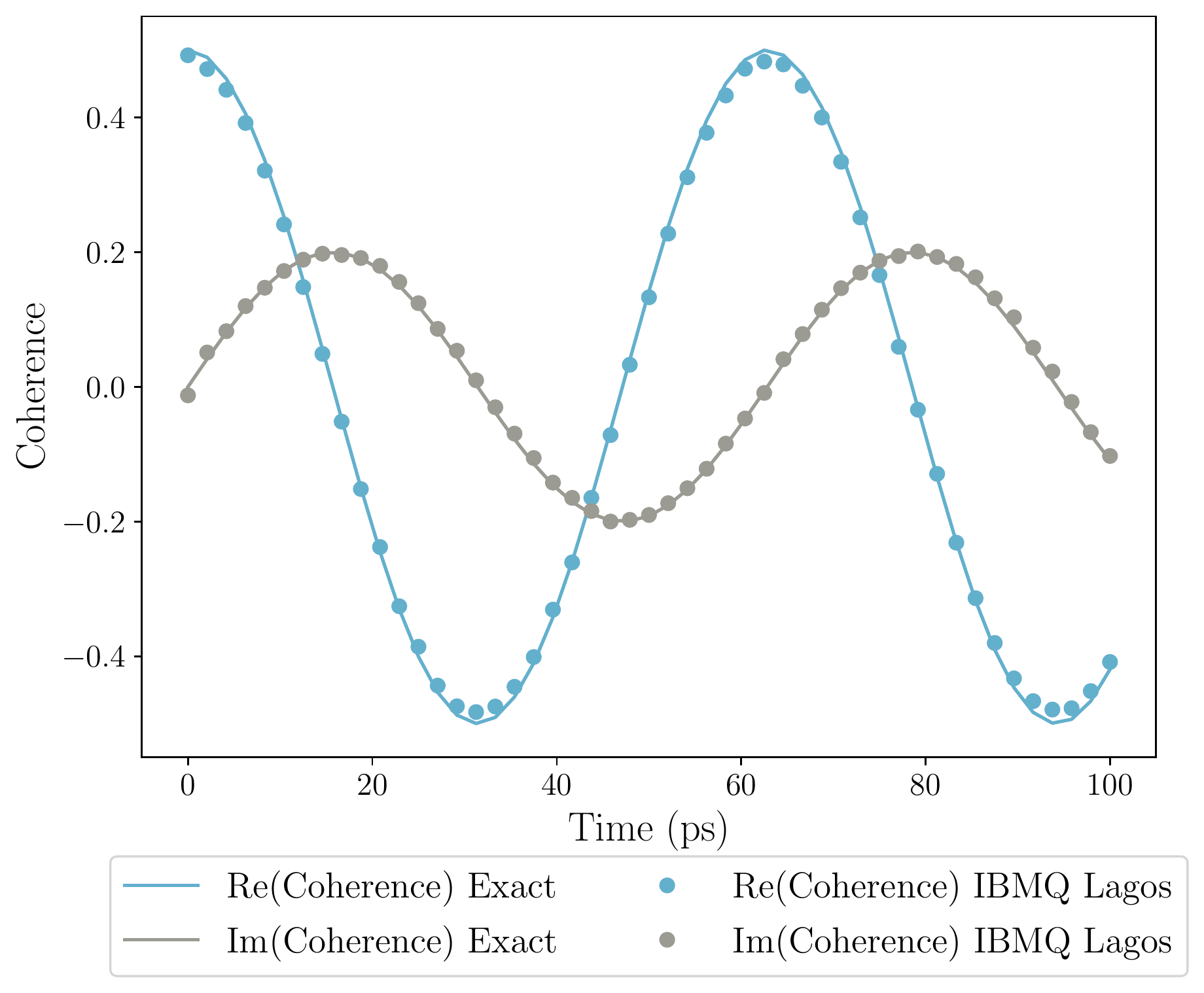}
    \caption{Real and imaginary parts of the off-diagonal element (coherence) of a two-level state in a $Z \otimes Z$ dephasing channel. The results from the quantum device (dots) are accurate for all times in the simulation, where the exact solution (lines) is generated classically. The results from the quantum computer are generated from density matrix tomography.}
    \label{fig:zzSDynamics}
\end{figure}
Figure ~\ref{fig:zzSDynamics} shows the time dynamics of the dephasing channel, with the exact classical solution (lines) and results from IBMQ Lagos (circles). The initial state is chosen as $\frac{1}{\sqrt{2}}(|0\rangle + |1\rangle)$, and the populations are not time dependent, so only the coherence is shown. We performed tomography of the density matrix to generate the coherences from the quantum device, and the simulation is very accurate over the time range.

\begin{figure}
    \centering
    \includegraphics[width=\columnwidth, trim={1in 1.2in 1in 1.5in}, clip]{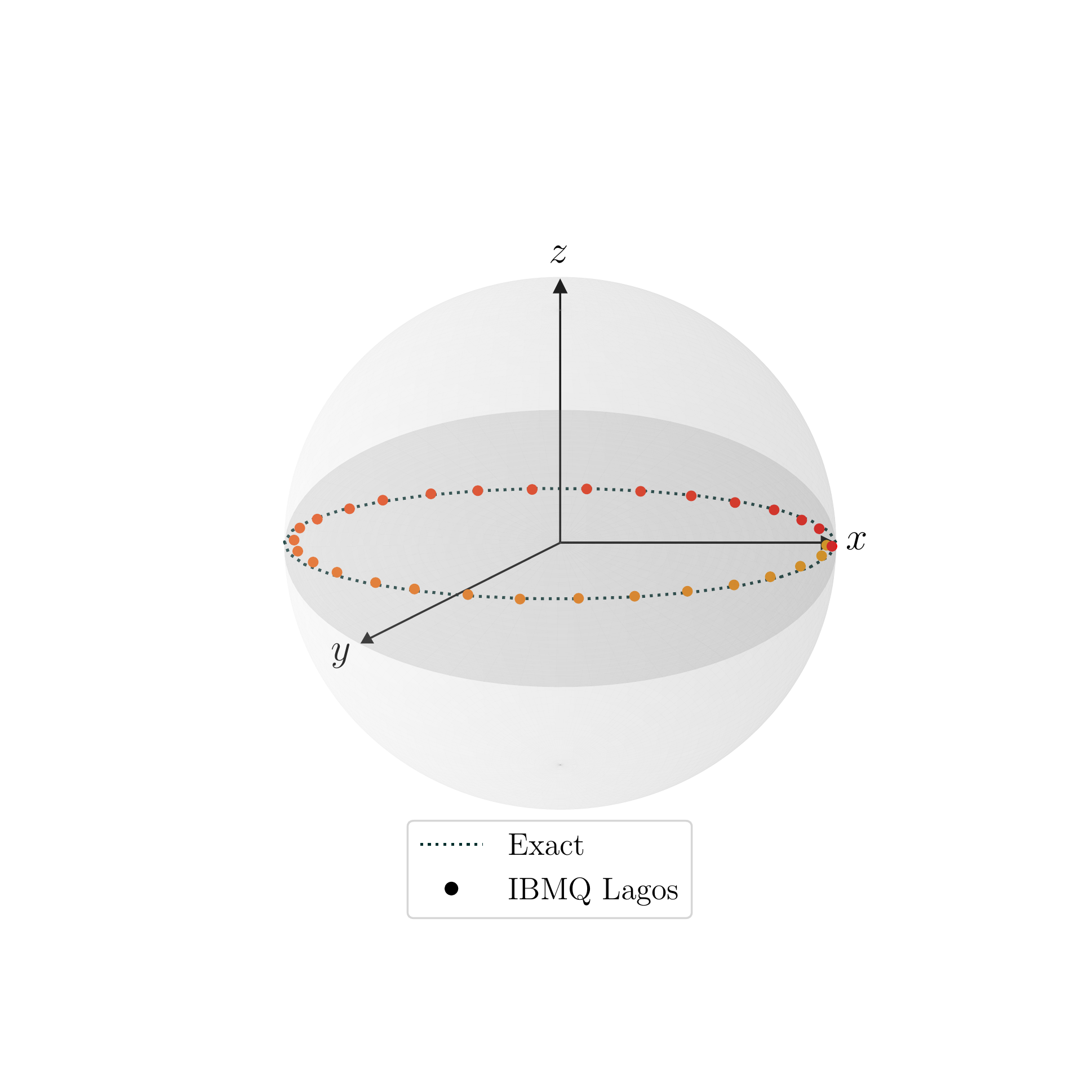}
    \caption{The trajectory of a two-level system  undergoing dephasing with a one-qubit environment is shown on a Bloch sphere with exact results represented by the dotted line, and results from IBMQ Lagos represented by dots. The trajectory is in the $z=0$ plane represented by the dark grey circle. One period is shown (about 60 ps), where the initial state, $\frac{1}{\sqrt{2}}(|0\rangle + |1\rangle)$, is the first red dot on the right. The system explores mixed states when it is in the interior of the sphere.}
    \label{fig:zzSBloch}
\end{figure}

Another way to visualize the time dynamics of this process is with the Bloch sphere. Because the populations are not time-dependent, the state stays in the $z=0$ plane of the Bloch sphere because of the choice of initial state. We show the Bloch sphere for the $Z \otimes Z$ dephasing process in Fig.~\ref{fig:zzSBloch}.
The dotted line is the exact trajectory of the state, and the dots are the simulated trajectory. Beginning in red with the initial state, we show one period of the recurring dephasing process, which is about 60 ps. The system is exploring the interior of the sphere, which indicates the system is a mixed state at those points.

For the second example, we simulate the time evolution of a two-level system in a zero-temperature amplitude damping channel. The Kraus operators in this case are,
\begin{equation}
\begin{aligned}
\hat K_0 &= \begin{pmatrix}1	& 0 \\ 0 &  \sqrt{e^{-\gamma t}} \end{pmatrix} =  I  \begin{pmatrix}1& 0 \\ 0 &  \sqrt{e^{-\gamma t}} \end{pmatrix}   I\\
\hat K_1 &= \begin{pmatrix} 0 & \sqrt{1-e^{-\gamma t}} \\ 0 & 0 \end{pmatrix} =  I
\begin{pmatrix}\sqrt{1-e^{-\gamma t}} & 0 \\ 0 & 0 \end{pmatrix} X,
\end{aligned}
\end{equation}
where we have emphasized the form of the SVD in the last equality. Here $I$ is the identity, $X$ is the Pauli-X operator, and $\gamma$ is the decay rate. We use $\gamma=0.15$ ps$^{-1}$ for the simulations here. The chosen initial state of the system is mixed and given by,
\begin{equation}
    \rho(0) = \frac{1}{4} \begin{pmatrix} 1 & 1 \\ 1 & 3  \end{pmatrix}.
\end{equation}
This state can be decomposed into $|\psi_0\rangle = |1\rangle$ and $|\psi_1\rangle= \frac{1}{\sqrt{2}}(|0\rangle + |1\rangle)$, which are prepared with an $X$ gate and $H$ gate, respectively.

\begin{figure}
    \centering
    \includegraphics[width=\columnwidth]{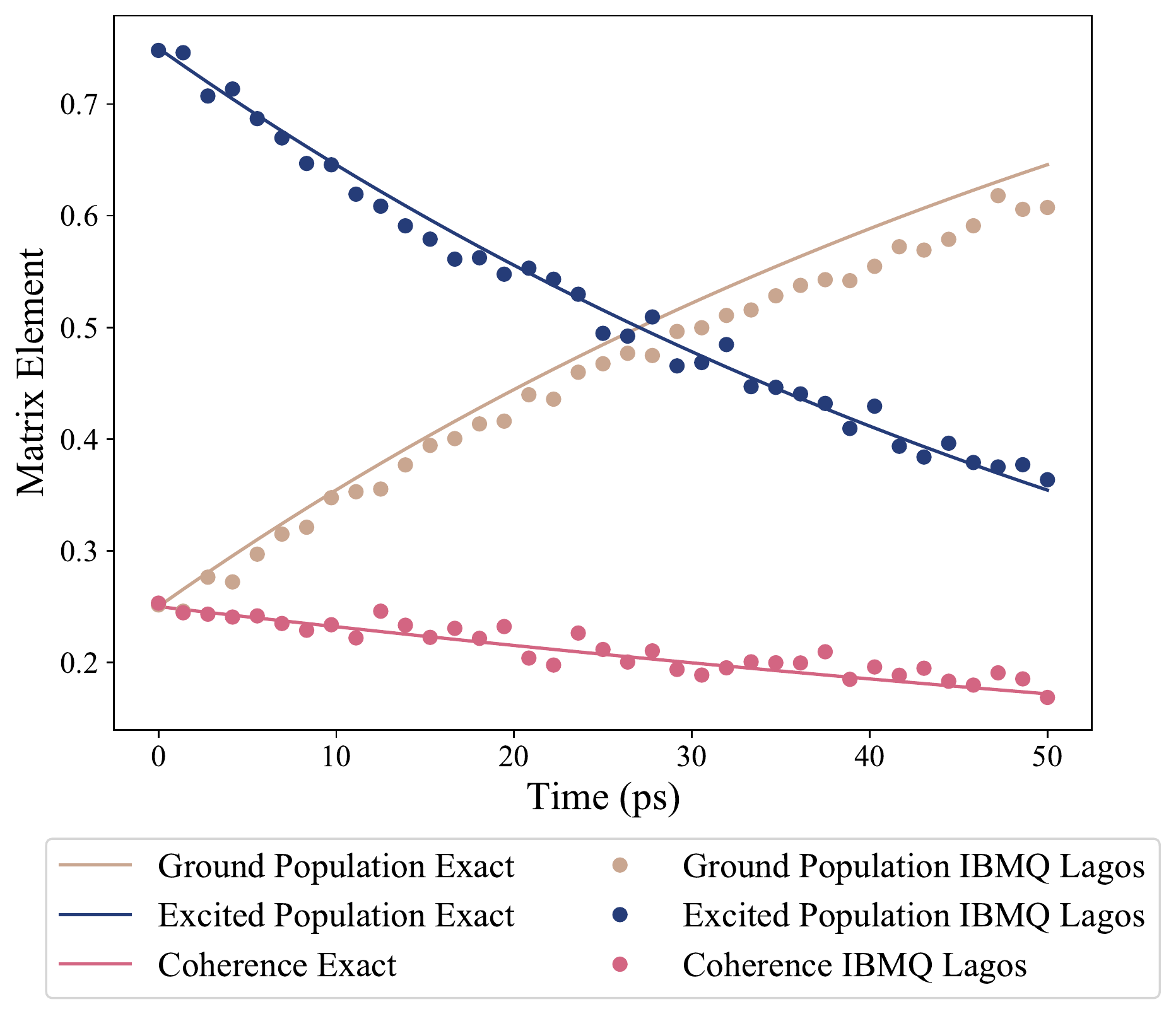}
    \caption{Time dynamics of a zero-temperature two-level system in an amplitude damping channel. The results from the IBMQ Lagos device (dots) are in good agreement with the exact results (solid lines). Results generated from a single run using the device maximum 32000 samples (shots) with error mitigation.}
    \label{fig:adamp}
\end{figure}

The results of the simulation of the two-level amplitude damping channel are shown in Fig.~\ref{fig:adamp}. The experimental results (dots) are in good agreement with the exact solution (solid lines). The populations and coherences are all modeled accurately, and demonstrate the non-unitary evolution of the system as the system loses energy over time.

Description of non-unitary evolution through singular-value decomposition of the Kraus maps allows for accurate simulation of open quantum systems on unitary-gate-based quantum computers. While the SVD has a classical cost if computed numerically, in the cases presented here, the SVD of the Kraus maps is immediately available by inspection.

\section{Discussion}

Implementation of non-unitary operations on quantum devices has important applications such as sub-normalized state preparation and non-unitary dynamics. Here we presented an algorithm based on diagonal non-unitary operators, which simplifies the implementation of dense non-unitary operators. Constrained by the limited qubit number and possible gate depths in the NISQ regime, many algorithms struggle with scaling for practical implementation, due to the expense of dense unitaries operating over the dilated Hilbert space.

Our algorithm implements a non-unitary operator with two unitaries on the original Hilbert space, and a diagonal operator on the dilated Hilbert space. We show that any non-unitary operator can be implemented in this way using the SVD. Assuming the classical cost of the SVD, and in the limit of large system size, the circuit complexity of our algorithm is approximately half as much as a dense dilated unitary. Seen another way, our algorithm results in circuits about twice as deep compared to a unitary on the original Hilbert space.

We achieve this implementation by realizing a diagonal unitary dilation for non-unitary diagonal operators. Any non-unitary diagonal operator can be dilated to a unitary diagonal operator with only one qubit, assuming the magnitude of each element is bounded above by one. Computing the elements of the dilated unitary is straightforward, and results in an exact representation of the diagonal operator. Unitary diagonals can be exactly implemented with $2^{d+1} - 3$ one- and two-qubit gates, for $d$ qubits; however, they can also be approximately implemented with polynomial gate scaling. Future work will include exploring how approximate implementation would effect the accuracy of the overall simulation of non-unitary processes.

Our algorithm is probabilistic in that it depends on the measured state of the ancilla qubit. When the ancilla is in state $|0\rangle$, the algorithm is successful, and we have exactly prepared the state after application of a non-unitary diagonal. Future work could include implementing the algorithm with amplitude amplification techniques, which improve the success probability of the probabilistic algorithm. This helps reduce the error from noise, or reduces the number of samples required for accurate simulations.

We discussed applications of this algorithm in the context of state preparation. While our probabilistic algorithm can be used to prepare normalized states, it is better suited for the preparation of sub-normalized states, which arise in the context of non-unitary state transformations. We demonstrated the preparation of one-qubit sub-normalized quantum states on IBM's quantum computer Lagos with high accuracy and fidelity using our algorithm, and show the states can be prepared accurately with a relatively modest number of samples or shots.

We also applied our algorithm to two time-dependent open quantum systems, which undergo non-unitary evolution. Using the SVD, we showed that any non-unitary operator can be decomposed and implemented with one ancilla qubit using the unitary diagonal construction for the singular values. We used the SVD to decompose the Kraus operators for the dynamics; however, the algorithm is applicable to any operator. We accurately reproduced the non-unitary dynamics of a two-level dephasing channel and a two-level amplitude damping channel using IBM's Lagos quantum computer. Our algorithm is particularly useful when the cost of the SVD is not prohibitive, which is the case for most operators considered in the NISQ era. Furthermore, because the diagonal unitary dilation is exact, we can implement the entire non-unitary process on one quantum circuit, which is useful when further transformation of the state is desired.

Our algorithm is effective in implementing non-unitary operations in quantum simulation in the NISQ era. Beyond state preparation and non-unitary dynamics, our algorithm will also be useful in computing expectation values of non-unitary observables without controlled unitaries in the Hadamard test. Quantum-classical hybrid methods may also benefit from our algorithm, when the SVD of the relevant operators can be computed. Our algorithm provides an intuitive look at the action of non-unitary operators in the qubit space, and reduces the dilation problem to the implementation of a dilated diagonal unitary based on the singular values of an operator.

\begin{acknowledgments}

This work is supported by the NSF RAISE-QAC-QSA, Grant No. DMR-2037783 and the Department of Energy, Office of Basic Energy Sciences Grant DE-SC0019215.  We acknowledge the use of IBM Quantum services for this work. The views expressed are those of the authors, and do not reflect the official policy or position of IBM or the IBM Quantum team.

\end{acknowledgments}

\bibliography{svd_oqs}

\end{document}